\documentclass[preprint]{aastex62}
\usepackage{amsmath,amstext}
\usepackage{graphicx}
\received{}
\revised{}
\accepted{}

\shortauthors{Shi et al.}
\begin{document}

\title{Forward Modeling of Simulated Transverse Oscillations in Coronal Loops and the Influence of Background Emission}

\correspondingauthor{Mijie Shi}
\email{mijie.shi@kuleuven.be}

\author{Mijie Shi}
\affiliation{Centre for mathematical Plasma Astrophysics, Department of Mathematics, KU Leuven, B-3001 Leuven, Belgium}
\affiliation{Shandong Key Laboratory of Optical Astronomy and Solar-Terrestrial Environment, Institute of Space Sciences, Shandong University, Weihai, Shandong, 264209, China}

\author{Tom Van Doorsselaere}
\affiliation{Centre for mathematical Plasma Astrophysics, Department of Mathematics, KU Leuven, B-3001 Leuven, Belgium}

\author{Patrick Antolin}
\affiliation{Department of Mathematics, Physics and Electrical Engineering, Northumbria University, Newcastle Upon Tyne, NE1 8ST, UK}

\author{Bo Li}
\affiliation{Shandong Key Laboratory of Optical Astronomy and Solar-Terrestrial Environment, Institute of Space Sciences, Shandong University, Weihai, Shandong, 264209, China}



\begin{abstract}
We simulate transverse oscillations in radiatively cooling coronal loops
and forward-model their spectroscopic and imaging signatures,
paying attention to the influence of background emission.
The transverse oscillations are driven at one footpoint by a periodic velocity driver. 
A standing kink wave is subsequently formed
and the loop cross-section is deformed due to the Kelvin-Helmholtz instability, resulting in energy dissipation and heating at small scales.
Besides the transverse motions,
a long-period longitudinal flow is also generated due to the ponderomotive force induced slow wave.
We then transform the simulated straight loop to a semi-torus loop
and forward-model their spectrometer and imaging emissions,
mimicking observations of Hinode/EIS and SDO/AIA.
We find that the oscillation amplitudes of the intensity are different at different slit positions,
but are roughly the same in different spectral lines or channels.
X-t diagrams of both the Doppler velocity and the Doppler width show periodic signals.
We also find that the background emission dramatically decreases the Doppler velocity,
making the estimated kinetic energy two orders of magnitude smaller than the real value.
Our results show that background subtraction can help recover the real oscillation velocity.
These results are helpful for further understanding transverse oscillations in coronal loops
and their observational signatures.
However, they
cast doubt on the spectroscopically estimated energy content of transverse waves using the Doppler velocity.
\end{abstract}




\section{Introduction} 
     \label{S-Introduction} 

Since their first imaging detections by the Transition Region and Coronal Explorer (TRACE)
\citep{1999ApJ...520..880A,1999Sci...285..862N},
transverse oscillations have been abundantly observed in coronal loops
\citep[e.g.,][]{2015A&A...577A...4Z,2016A&A...585A.137G}.
This kind of transverse oscillations, with large amplitude but strong damping,
are usually associated with solar flares or coronal mass ejections 
\citep{2002SoPh..206...99A}.
Recent observations also see another kind of small-amplitude transverse oscillations showing weak or no decay
\citep{2012ApJ...759..144T,2012ApJ...751L..27W}.
These decayless oscillations are not related to solar eruptive events and are also frequently observed in coronal loops
\citep[e.g.][]{2013A&A...552A..57N,2013A&A...560A.107A,2015A&A...583A.136A}.
The generation mechanism of decayless oscillations is still under debate.
\cite{2016ApJ...830L..22A} showed that the decayless oscillations can be a result of the combined effects of periodic brightenings and the coherent motions of the Kelvin-Helmholtz instability (KHI) vortices.
\cite{2016A&A...591L...5N} proposed that 
decayless oscillations can be generated by a self-oscillatory process, which was recently achieved in simulations
by \cite{2020ApJ...897L..35K,2021ApJ...908L...7K}.
Decayless oscillations were also produced
in recent simulations by implementing velocity drivers at the loop footpoint
\citep{2017A&A...604A.130K,2019ApJ...870...55G,2020A&A...633L...8A}.

Transverse oscillations in coronal loops are often attributed to kink waves supported by the structured solar corona
\citep{1983SoPh...88..179E}.
Energy carried by kink waves can convert or dissipate through resonant absorption \citep{1988JGR....93.5423H,2002A&A...394L..39G,2002ApJ...577..475R,2012A&A...539A..37P}, 
phase mixing 
\citep{1983A&A...117..220H,2017A&A...601A.107P},
or Kelvin-Helmholtz instability \citep{1984A&A...131..283B,
	2008ApJ...687L.115T,2015ApJ...809...72A},
and thus could have potential heating effects to the solar corona.
A number of three-dimensional (3D) simulations have shown the heating effects of footpoint driven kink waves
\citep{2017A&A...604A.130K,2019A&A...623A..53K,2019ApJ...870...55G,2019ApJ...883...20G,2019ApJ...876..100A}.
Recent simulations by \cite{2021ApJ...908..233S} showed that the energy dissipation of kink waves is a promising mechanism for balancing the radiative losses of coronal loops.

Radiative cooling directly decreases the plasma temperature and thus influences the extreme ultraviolet (EUV) emissions.
\cite{2008ApJ...686L.127A}
concluded that in order to avoid contamination from other loops,
the oscillation of a single loop can not be detected for longer than 10-20 mins in one passband.
Radiative cooling of coronal loops can also influence the kink wave properties and 
the period ratio between the fundamental and first harmonic kink waves
\citep{2009ApJ...707..750M}.
\cite{2011SoPh..271...41R,2011A&A...534A..78R}
showed that cooling can cause amplification of kink waves but this amplification is inefficient compared with resonant damping.
This ineffective amplification caused by radiative cooling was later shown in simulations by
\cite{2015A&A...582A.117M}.

Forward modeling is an approach that synthesizes plasma emissions and converts numerical models to observables.
A variety of forward modeling analysis were conducted
exploring the synthetic emissions modulated by magnetohydrodynamic (MHD) waves in coronal loops,
such as slow waves
\citep{2015ApJ...807...98Y,2016ApJ...820...13M},
kink waves
\citep{2016ApJS..223...23Y},
and sausage waves
\citep{2013A&A...555A..74A,2019ApJ...883..196S}.
\cite{2014ApJ...787L..22A,2017ApJ...836..219A}
investigated the observational signatures of fine structures and dynamic instabilities induced by transverse oscillations. 

The Doppler velocity is one of the key parameters to estimate  wave energy from real observations.
\cite{2007Sci...317.1192T} found the observed wave energy in the line of sight (LOS) is not enough to heat the corona.
However, some energy could be hidden from the Doppler velocity as the consequence of the LOS superposition
\citep{2012ApJ...761..138M,2019ApJ...881...95P,2020ApJ...899....1P}.
\cite{2012ApJ...746...31D} showed that in the case of a multistrand loop the observed kinetic energy using the Doppler velocity could underestimate the actual energy by one or two orders of magnitude.
But in that paper, they simply integrated the velocity component along a LOS to estimate the kinetic energy,
which could be very different from the Doppler velocity obtained through forward modeling.
\cite{2017ApJ...836..219A} forward-modeled transverse MHD waves excited impulsively and studied the effect from dynamic instabilities associated with these waves. They found that the effect of resonant absorption and Kelvin-Helmholtz instability reduces the available kinetic energy by 90\%.
A doubling of the periodicity in the line width was also found. 
Furthermore, they observed a clear increase of the non-thermal line widths, but it did not compensate for the apparent loss of kinetic energy from the Doppler velocities.

In this work, we simulate transverse oscillations in  coronal loops using 3D MHD simulations 
and forward-model the synthetic observations
targeting at
the EUV Imaging Spectrometer (EIS) onboard Hinode
and the Atmospheric Imaging Assembly (AIA)
onboard Solar Dynamic Observatory (SDO).
While \cite{2016ApJS..223...23Y} forward-modeled the kink wave model in a semi-torus loop,
the difference of our work is that we
use real simulation data with the radiative cooling
 explicitly included.
We also study the influence of background emission by changing the integration column depth.
Section \ref{S-setup} is the simulation setup.
Sections \ref{S-results} and \ref{S-fomo} 
show the results of simulation and
forward modeling.
Section \ref{S-summary} is the summary and conclusion.

\section{Numerical Model}
\label{S-setup}
In our model, we set the coronal loop as a straight, density enhanced magnetic flux tube along the Cartesian $z$ direction.
Gravity is ignored.
The initial loop parameters depend only on the transverse position $x$ and $y$.
The density is given by,
\begin{eqnarray}
	\rho = \rho_e + (\rho_i - \rho_e)\zeta(x,y),\quad
	\zeta(x,y) = \frac{1}{2}(1-{\rm{tanh}}(b(\sqrt{x^2+y^2}/R-1))),
\end{eqnarray}
where $[\rho_i,\rho_e] = [2\times10^8,10^8]m_p~{\rm cm^{-3}}$   are the
densities in the interior and exterior of the loop
and $m_p$ is the proton mass.
$R = 1~\rm{Mm}$ is the loop radius.
 The density changes smoothly from $\rho_i$ to $\rho_e$ 
within a transition layer $l$ determined by $b$.
We choose $b=20$, resulting in an $l\approx 0.3R$.
The initial temperature is $T = 1~\rm{MK}$ across the whole domain.
The magnetic field is along the $z$ direction,
with its strength inside the loop being 
$B_i=20~\rm{G}$.
The other initial parameters are used to maintain total pressure balance.

Similar to our previous work \citep{2021ApJ...908..233S},
we generate kink waves by implementing a 
dipole-like transverse velocity driver at one footpoint \citep{2010ApJ...711..990P}. 
Inside the loop the velocity is given by
\begin{eqnarray}
	[v_x,v_y]=[v_0{\rm{cos}}(\frac{2\pi t}{P_k}),0].
\end{eqnarray}
Outside the loop the velocity is 
\begin{eqnarray}
	[v_x,v_y]=v_0{\rm{cos}}(\frac{2\pi t}{P_k})R^2[\frac{x^2-y^2}{(x^2+y^2)^2},\frac{2xy}{(x^2+y^2)^2}].
\end{eqnarray}
$P_k$ is the period of the fundamental kink mode.
In our model the loop length $L = 200~\rm{Mm}$, 
so that $P_k = 2L/c_k = 112~\rm{s}$, 
where $c_k = 3562~\rm{km~s^{-1}}$ is the kink speed.
$v_0 = 8~\rm{km/s}$ is the amplitude of the velocity driver.

The simulations are performed using the PLUTO code 
\citep{2007ApJS..170..228M}
with radiative cooling taken into account.
The radiative loss function is from the CHIANTI database
\citep{2019ApJS..241...22D}.
Physical dissipation and thermal conduction are not included in our model. 
The simulation domain is
$[-10,10]~\rm{Mm}$, $[-10,10]~\rm{Mm}$, and $[0,200]~\rm{Mm}$
in $x$, $y$, and $z$ directions.
There are 800 uniform grid points in both the $x$ and $y$ directions,
and 100 uniform grid points in the $z$ direction, corresponding to a spatial resolution of $25~\rm{km}$ in the $x$ or $y$ direction.
We use outflow boundary conditions
for all quantities at the side boundaries.
The transverse velocities at the bottom boundary ($z=0$) are given by the driver,
and at the top boundary are fixed to be zero.
At both the top and bottom boundaries,
the $z$ component of velocity is antisymmetric.

\section{Simulation Results} %%%%%%%%%%%%%%%%%%%%%%%%%%%%%%%%%%%%%%%%
\label{S-results}     

\begin{figure}    %%%%%%%%%%%%%%%%%% FIGURE 1
\begin{center}
	\includegraphics[width=0.8\textwidth,clip=]{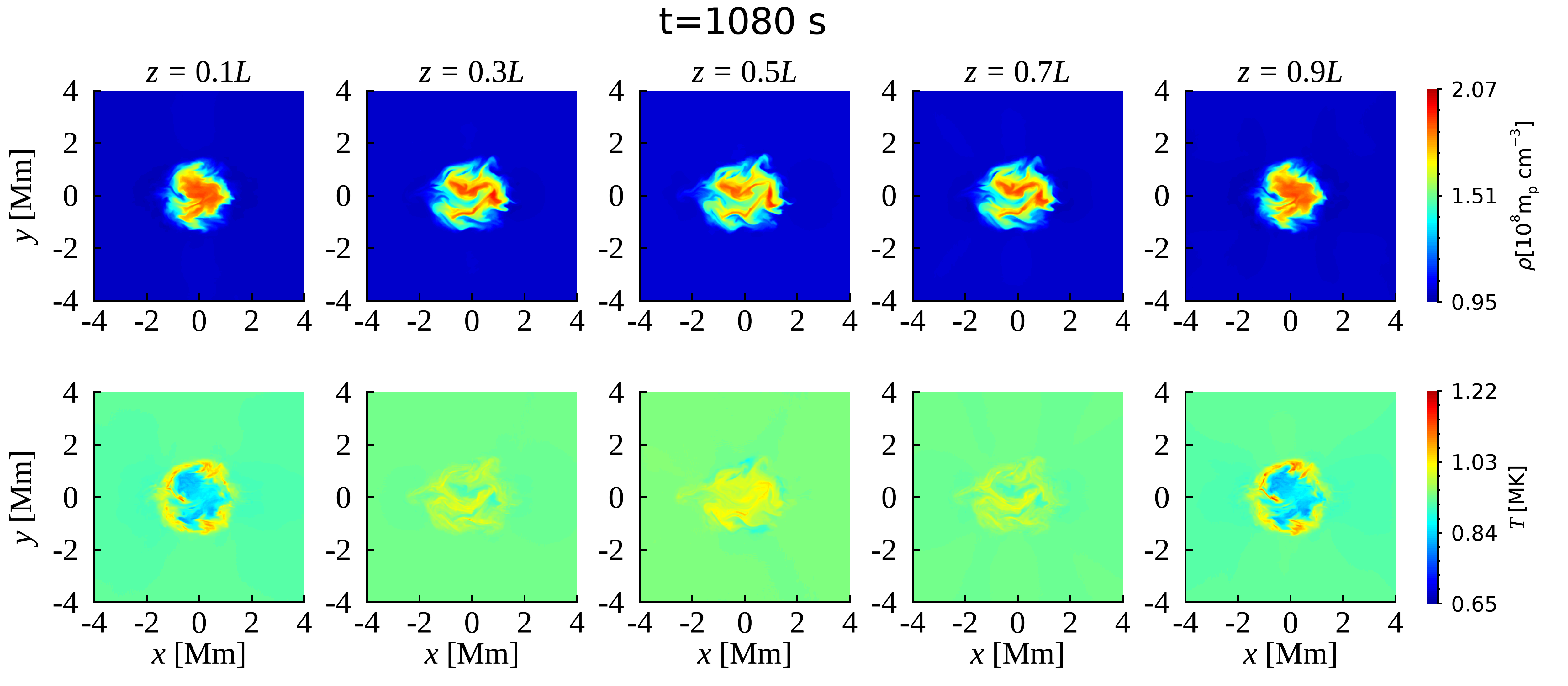}
		\caption{Density $\rho$ and temperature $T$ profiles at the loop cross section of five $z$ locations (from $0.1L$ to $0.9L$) at $t = 1080 $ s ($\approx 9.6P_k$).
	An animated version of this figure is available. The animation has the same layout as the static figure, and runs from 0 -- 2400 s.
	\newline
	(An animation of this figure is available.)	
}
	\label{F-cross_section}
\end{center}
\end{figure}

The velocity driver implemented at one footpoint generates a kink wave that propagates along the loop and reflects at the other footpoint.
Because the driver period is the same as that of the fundamental kink wave,
a standing kink wave forms after several periods,
and at the same time the Kelvin-Helmholtz instability (KHI) develops due to the velocity shear.
Figure \ref{F-cross_section} shows the density and temperature distributions in five different cross sections 
at $t=1080$ s ($\approx 9.6P_k$).
The profiles at $z = 0.1L$ and $0.9L$ (also at $z = 0.3L$ and $0.7L$) are almost identical to each other,
in line with the fundamental standing kink wave.
From the associated animation, we see that KHI develops after $2P_k$ and later deforms the loop cross sections to turbulent structures,
with both the KHI and loop deformation strongest at the apex ($z = 0.5L$).
The radiative cooling is proportional to density squared
and thus has the most effect on high density regions.
This is shown in Figure \ref*{F-cross_section} near the footpoint, and one sees that the temperature in the high-density interior is lower than in the loop exterior.
As small scale structures are generated at a later stage,
numerical dissipation will occur and heat the plasma.
The estimated numerical resistivity and viscosity are $1.7\times10^{-9}~\rm{s}$ and $8.6\times10^{-5}~\rm{g ~(cm~s)^{-1}}$, respectively, by taking into account energy conservations \citep{2021ApJ...908..233S}.
The corresponding (magnetic) Reynolds number is on the order of $10^{5}$.
We see that the temperature distributions around the footpoint 
and at the apex are a bit different. 
Around the footpoint, we see a cooler core region surrounded 
by a hotter ring.
This is a combined effect from both heating and cooling,
as dissipation and heating are stronger around the loop edges, 
and radiative cooling is stronger in the loop core.  
At the apex, where the loop is more turbulent,
temperatures at this cross-section are effectively equilibrated.

\begin{figure}    %%%%%%%%%%%%%%%%%% FIGURE 2
\begin{center}
	\includegraphics[width=0.8\textwidth,clip=]{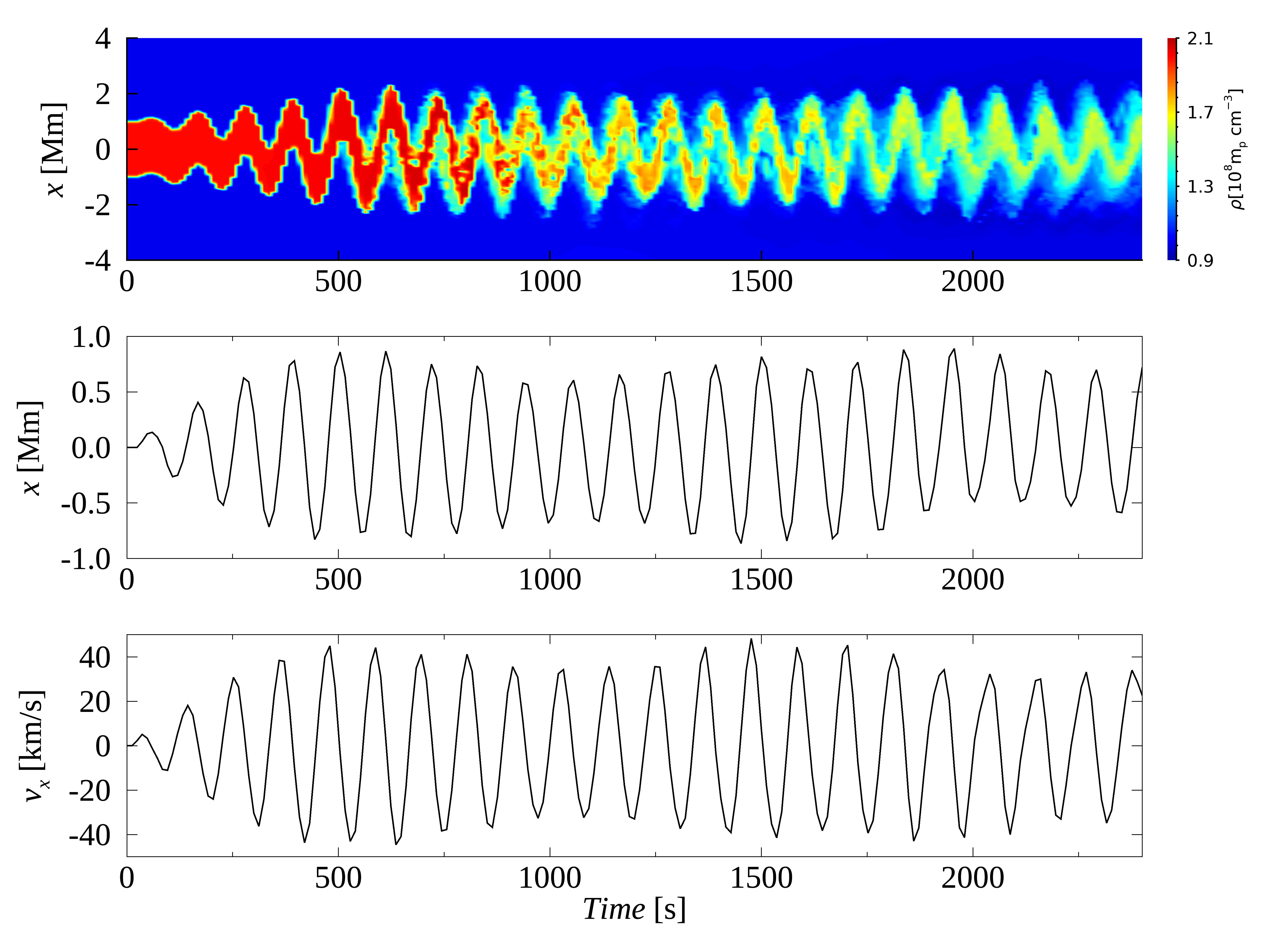}
	
	\caption{Top: time-distance map of density at the mid-plane (a slice at $y = 0$) of the loop apex ($z = 0.5L$).
		Middle: oscillation amplitude from calculating the centroid of the loop interior ($\rho > 1.1\rho_e$) at the apex.
		Bottom: oscillation velocity calculated from the Middle panel.}
	\label{F-kink_oscil}
	\end{center}
\end{figure}

The oscillation amplitude and velocity of kink wave are shown in Figure \ref{F-kink_oscil}.
The top panel of Figure \ref{F-kink_oscil} is the time-distance map of density at the apex. 
The oscillation amplitude grows first and then saturates after several periods.
Despite the deformation of the loop cross section and the generation of small scale structures,
the loop oscillation is still clearly recognized with no obvious decay,
which is similar to the observed decayless oscillations
as reported by \cite{2019FrASS...6...38K}.
The oscillation amplitude and velocity shown in the middle and bottom panels of Figure \ref{F-kink_oscil} are calculated by
tracking the centroid of the loop interior regions ($\rho > 1.1\rho_e$).
From Figure \ref{F-kink_oscil}, we find that the oscillation amplitudes look larger than typical values of decayless oscillation observed in coronal loops
\citep{2013A&A...552A..57N,2013A&A...560A.107A},
also the oscillation velocities are larger than those of kink waves observed from a spectrometer
\citep{2008A&A...487L..17V}.
However, it is inappropriate to directly compare our model with the observed ones,
because both the observed oscillation amplitudes and velocities strongly depend on the angle between the LOS and the oscillation axis.
The observed Doppler velocities,
which are derived from the emission integral along a LOS,
are often smaller than the real oscillation velocities,
as shown by \cite{2017ApJ...836..219A}.
Also, the low spatial and spectral resolutions can decrease the observed velocities. 

\begin{figure}   %%%%%%%%%%%%%%%%%% FIGURE 3
	\centerline{\includegraphics[width=0.6\textwidth,clip=]{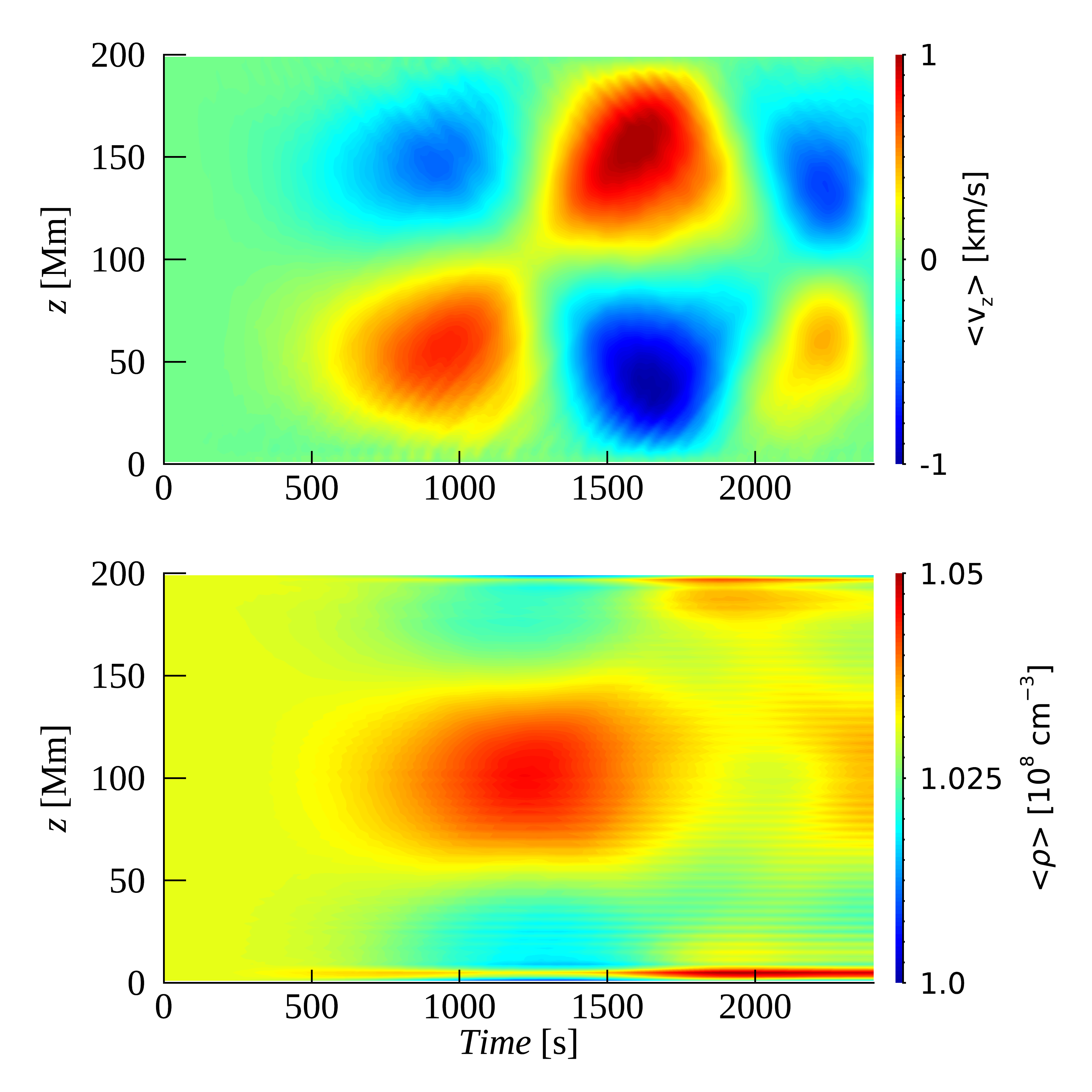}
	}
	\caption{Cross section ($|x|,|y| \le 5$ Mm) averaged longitudinal velocity ($v_z$, top) and density ($\rho$, bottom).}
	\label{F-z_flow}
\end{figure}

In addition to the transverse oscillations,
we also see a long-period longitudinal flow in the $z$ direction.
The top panel of Figure \ref{F-z_flow} shows the evolution of the longitudinal velocity $v_z$ averaged in the $x-y$ plane ($|x|,|y| \le 5$ Mm).
Clearly we see a periodic profile,
which was also found in previous simulations
\citep{2016A&A...595A..81M,2017A&A...604A.130K,2019ApJ...883...20G}.
This longitudinal flow is caused by the slow magnetoacoustic wave driven by the ponderomotive force of large-amplitude standing wave,
which was previously discussed by
\cite{1994JGR....9921291R} and \cite{2004ApJ...610..523T}.
The analytical results from \cite{1994JGR....9921291R}
show that the frequency of this slow wave is
$\Omega = 2k_zc_s$, where $k_z$ is wavevector and $c_s$ is sound speed.
In our model, $k_z = \pi/L$ for fundamental wave, $c_s = 166$ km/s for 1 MK plasma,
so that the period is $2\pi/\Omega = L/ c_s = 1205$ s,
which shows good agreement with the period of $v_z$ in Figure \ref{F-z_flow}.
The longitudinal flow becomes weaker at the later stage ($t \approx$ 2200 s) probably due to the development of turbulent structures.
The consequence of this longitudinal flow is the mass enhancement at the apex, shown at the bottom panel of Figure \ref{F-z_flow}.
\cite{2004ApJ...610..523T} 
proposed that this mechanism could explain the observed emission enhancement at the apex of some oscillating loops,
though other physical processes, such as gravity, need to be considered.

\begin{figure}    %%%%%%%%%%%%%%%%%% FIGURE 4
	\centerline{\includegraphics[width=0.6\textwidth,clip=]{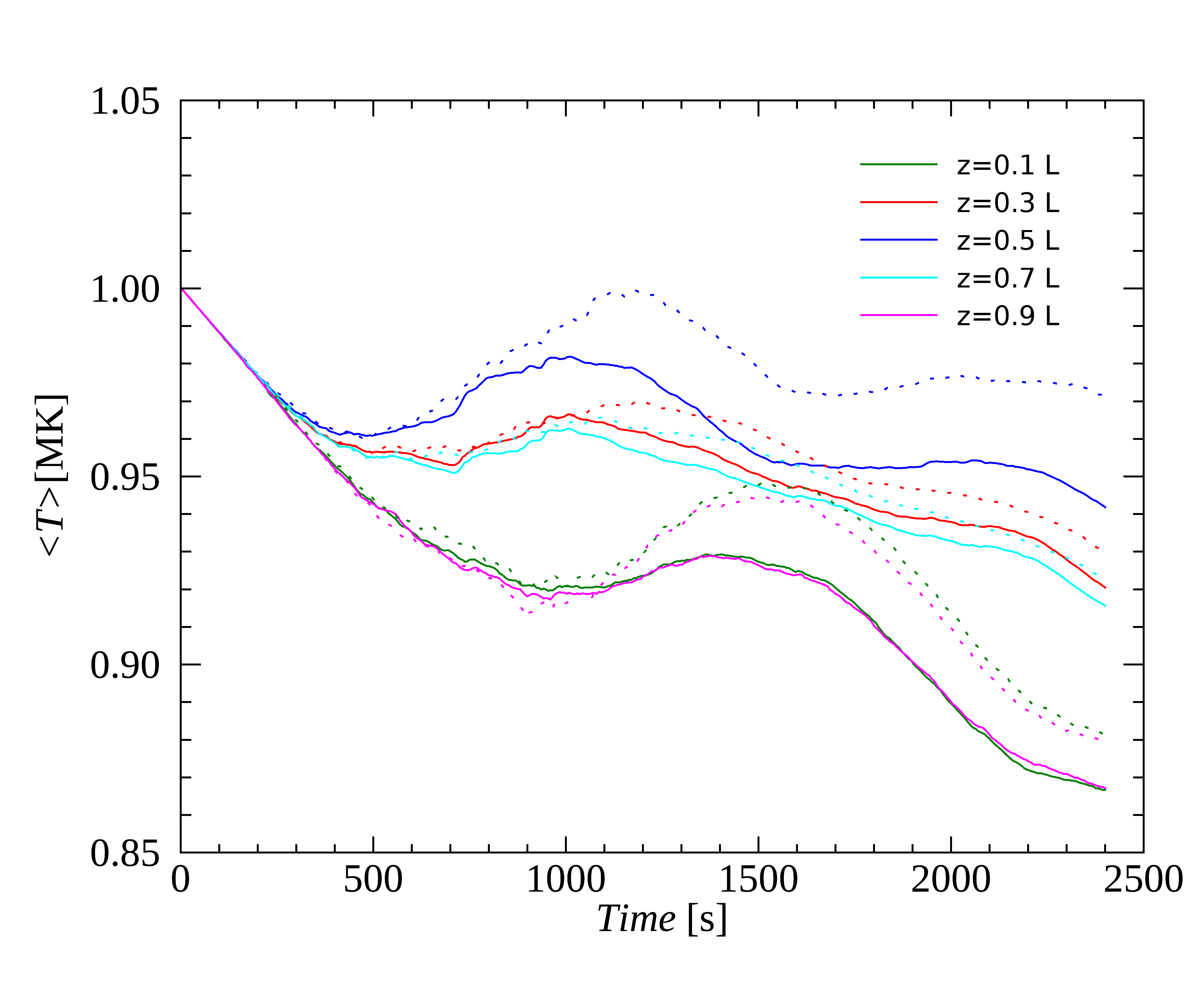}
	}
	\caption{Solid lines: volume-averaged temperature in the loop interior ($\rho \ge 1.1\rho_e$) at different $z$ cross sections ($z = 0.1L$ to $z=0.9L$).
	Dotted lines: same as the solid lines but for the case with coarse resolution.}
	\label{F-tem}
\end{figure}

Solid lines of Figure \ref{F-tem} show the averaged temperature of the loop interior ($\rho \ge 1.1 \rho_e$) at five cross sections.
On the whole we find a decreasing trend of the averaged temperature,
meaning that radiative cooling plays a dominant role.
The long-period modulation of the temperature profiles,
obviously seen at the apex ($z = 0.5L$) and 
around the footpoints ($z = 0.1L, 0.9L$),
are due to the ponderomotive force induced slow wave.
A similar temperature profile was also shown by
\cite{2017A&A...604A.130K}.
Looking at Figure \ref{F-tem}, we see that at the initial stage ($t<200$ s) all curves are overlapped,
but at the later stage temperatures at different cross sections are different,
with the temperature at the apex being higher than that around the footpoint.
This is because around the footpoint the cross section is less turbulent,
and thus the radiative cooling is stronger there
than at the apex.
Numerical dissipation will also play a role in heating the plasma,
and consequently counteract the radiative losses,
leading to longer cooling times.
Numerical dissipation is widely used in 3D simulations,
as the realistic (magnetic) Reynolds number in solar corona can not be achieved.
To get a sense of how numerical dissipation changes at different spatial resolutions,
we run an additional simulation case with the resolution 2 times coarser,
i.e., 400 grids in both $x$ and $y$ directions.
The dotted lines in Figure \ref{F-tem} show the temperatures in this case.
We find that the temperature is higher in the case of coarser resolution,
indicating a higher numerical dissipation. 

\section{Forward Modeling} %%%%%%%%%%%%%%
\label{S-fomo}
In this section we forward-model the synthetic emissions of a semi-torus loop illustrated in Figure \ref{F-los}.
This semi-torus loop, with zero inclination between the loop plane and the vertical to the solar surface,
 is obtained by simply bending the simulated straight loop.
Following \cite{2016ApJS..223...23Y},
the coordinate transformation from the straight loop ($x,y,z$)
to the semi-torus loop ($x_1,y_1,z_1$) is given by:
\begin{eqnarray}
	x_1 = x, ~
	y_1 = (R+y)\mathrm{sin}\zeta, ~
	z_1 = -(R+y)\mathrm{cos}\zeta,
\end{eqnarray}
where $\zeta = z/R$, and $R = L/\pi$.
The velocities are transformed by:
\begin{eqnarray}
	v_{x1} = v_x, ~
	v_{y1} = v_y\mathrm{sin}\zeta + v_z\mathrm{cos}\zeta, ~
	v_{z1} = -v_y\mathrm{cos}\zeta + v_z\mathrm{sin}\zeta.
\end{eqnarray}
In the new coordinates, the kink wave is horizontally polarized 
and oscillates perpendicular to the $y_1-z_1$ plane.

We forward-model the synthetic emissions using the FoMo code\footnote{https://wiki.esat.kuleuven.be/FoMo/}.
FoMo can generate both spectrometer and imaging observations.
For spectrometer observations like Hinode/EIS, 
FoMo calculates the monochromatic emissivity based on the contribution function of the corresponding spectral line and the LOS velocity.
For imaging observations like SDO/AIA,
FoMo also takes into account the response function of the corresponding channel.
The comprehensive description of FoMo was provided by
\cite{2016FrASS...3....4V}.

We select an oblique LOS (shown in Figure \ref{F-los}) for the forward modeling analysis.
For a given LOS, FoMo rotates to a new coordinate system determined by the LOS ($z'$) and the plane of sky (POS, i.e., $x_{\rm{pos}}-y_{\rm{pos}}$),
calculates the emissivity at each pixel in the new coordinates,
and integrates them along the LOS ($z'$).
For those pixels outside the simulation domain,
we specify a constant emissivity value 
which
equals the values at the loop exterior of the initial state.
By doing this, we assume that the emissions outside the simulation domain are constant and not evolving.
Furthermore,
we make sure that all the rays of this LOS have the same column depth of 118 Mm, i.e., the maximum range of $z'$.

Forward modeling is conducted with a pixel size of 0.1$''$
and, for the spectroscopic mode, a wavelength spacing of 3.3 km/s.
For spectrometer observations,
we degrade the spatial resolution to 3$''$
and the spectral resolution to 36 km/s,
similar to
EIS \citep{2007SoPh..243...19C}.
For imaging observations,
we degrade the spatial resolution to 1.5$''$,
similar to AIA \citep{2012SoPh..275...17L}.
The temporal cadence is 12 s for both the spectrometer and imaging analysis.
A comprehensive analysis of the influence of different resolutions was given by
\cite{2017ApJ...836..219A}.

\subsection{Spectrometer Observations}
For spectrometer observations,
we forward-model two emission lines, i.e.,
Fe XII 195.119 \AA~and Fe VIII 185.213 \AA~lines.
We select these two lines because both are strong emission lines observed by EIS \citep{2007PASJ...59S.857Y} 
and are often used in real observations.
The formation temperatures of the two lines are
log(T) = 6.2 and 5.7, respectively,
corresponding to higher and lower temperatures comparing with the initial temperature of 1 MK.

\subsubsection{Plane of Sky Images}
\begin{figure}    %%%%%%%%%%%%%%%%%% FIGURE 6
	\centerline{\includegraphics[width=0.8\textwidth,clip=]{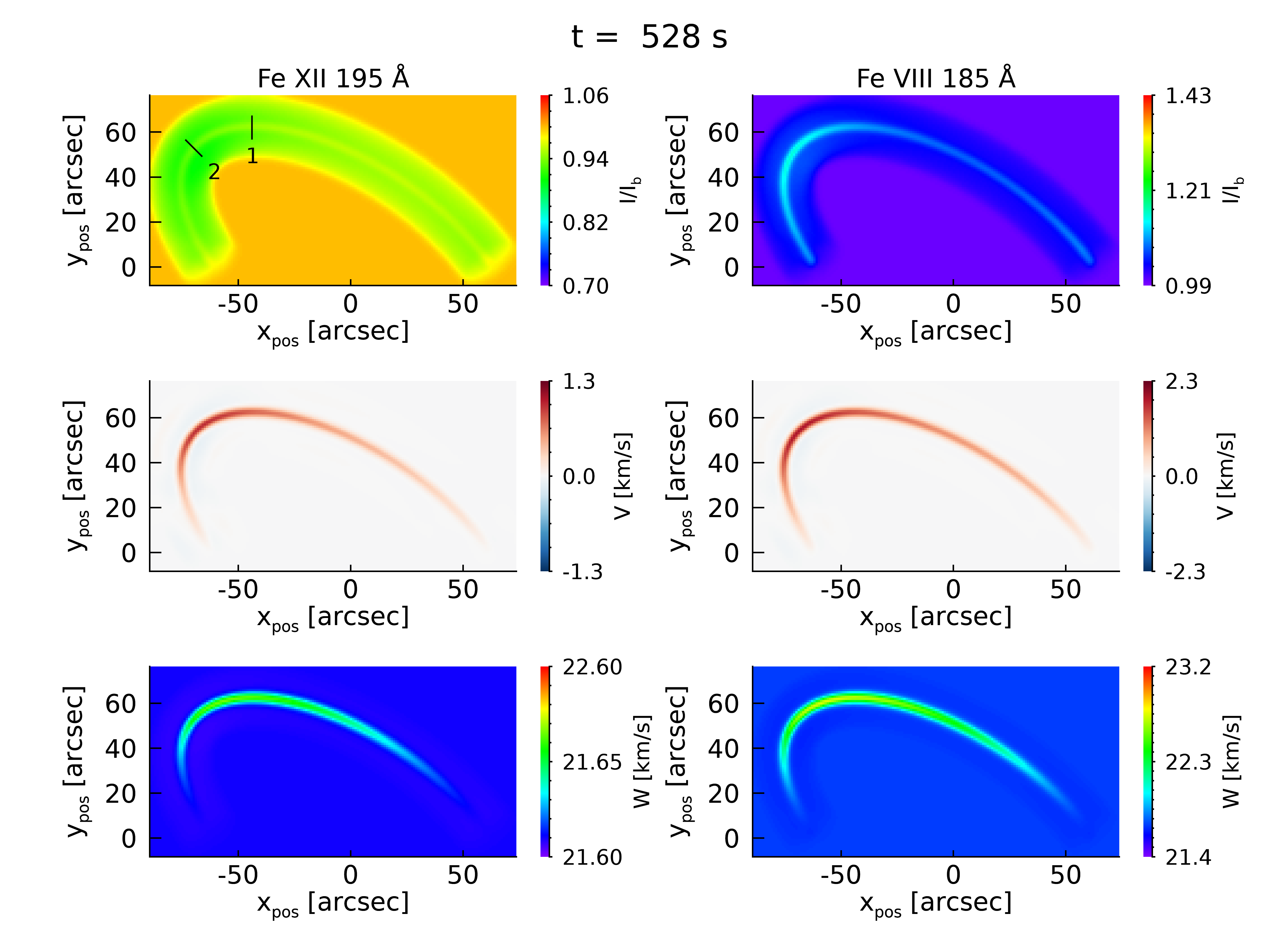}
	}
	\caption{Plane of sky images of normalized intensity ($I/I_b$),
		Doppler velocity ($V$), and Doppler width ($W$) for (left) Fe XII 195 \AA~and (right) Fe VIII 185 \AA~lines at $t = 528$ s.
		The black lines in the top-left panel mark the positions of two slits.
	An animated version of this figure is available. The animation has the same layout as the static figure, and runs from 0 -- 2400 s.
	\newline
	(An animation of this figure is available.)	
}
	\label{F-spec}
\end{figure}

At each pixel in the plane of sky,
we fit the spectral profile with a Gaussian
and get the intensity, Doppler velocity, and Doppler width.
Figure \ref{F-spec} shows the results at one snapshot.
The intensity is normalized by the background value in the plane of sky.
Here the background means the region where LOS rays do not 
pass through the simulation domain.
From the intensity images in Figure \ref{F-spec},
we can recognize the unchanging background and 
the region where LOS rays pass through the simulation domain.
The loop interior is also clearly seen.
In the simulation domain,
the intensity decreases for Fe XII 195 \AA~but increases for Fe VIII 185 \AA~because of their different formation temperatures.
The Doppler velocity is stronger at the loop interior.
The Doppler width is larger at the locations with larger velocity,
indicating the line broadening is dominated by the flow velocity.

\subsubsection{X-t Images}
Two slits marked at the top-left panel of Figure \ref{F-spec}
are selected in order to show the slit-cut results.
The center of slit 1 corresponds to the LOS ray passing through $[x,y,z] = [0,0,100]$ Mm (apex center) of the straight loop,
while the center of slit 2 corresponds to the ray passing through $[x,y,z] = [0,0,65]$ Mm.
The lengths of both slits are around 11$''$.
Slit-cut values are obtained by interpolating the plane of sky images and maintaining the pixel size.
Figure \ref{F-spec_slit_195} and Figure \ref{F-spec_slit_185} show the x-t images at both slits for Fe XII 195 \AA~and Fe VIII 185 \AA,
respectively.
At slit 1 the intensity oscillations are obvious for both lines.
But at slit 2 the oscillation amplitudes are small and 
not that obvious. 
Despite the opposite trends of the intensity variations for Fe XII 195 \AA~and Fe VIII 185 \AA,
we find that during our simulation time the oscillation amplitudes are roughly the same 
in the same slit.
At both slits, the Doppler velocity shows periodic blue and red shifts as the results of the kink oscillations.
The Doppler width also shows periodic enhancement,
with its period being only half of that of the Doppler velocity.
This is a similar result as \cite{2017ApJ...836..219A}.

\subsubsection{Influence of the Background Emission}
	In real observations, the emission column depth is an important value to derive some physical parameters such as the filling factor of a loop.
	However, the column depth can change as the loop's inclination or orientation changes
	\citep[e.g.,][]{2003A&A...397..765C,2010ApJ...717..458V,2012A&A...537A..49W}.
	For the loops possessing a low density contrast, like the one in our model,
	the emission contribution from the background is not negligible.
	Similarly, the column depth of the background could also change at a different LOS angle.
	Furthermore, 
	when observing a loop located at different locations,
	e.g., the disk or the limb,
	the background column depth would also be different.
	Thus it is interesting to study the 
	influence of different column depth of the background emission.
The background emission can influence the LOS integrated spectral profile and thus the Gaussian fitting \citep{2020A&A...634A..54P}.
We briefly discuss how the background emission influences our results by changing the column depth of the forward modeling analysis.
Besides the 118 Mm column depth used in the previous sections,
in this section we also adopt shorter column depths.
For each LOS ray, a shorter column depth is obtained by reducing the number of pixels
outside the simulation domain,
and thus reducing
the emission contribution from the unchanging background.

\begin{figure}    %%%%%%%%%%%%%%%%%% FIGURE 7
	\centerline{\includegraphics[width=0.5\textwidth,clip=]{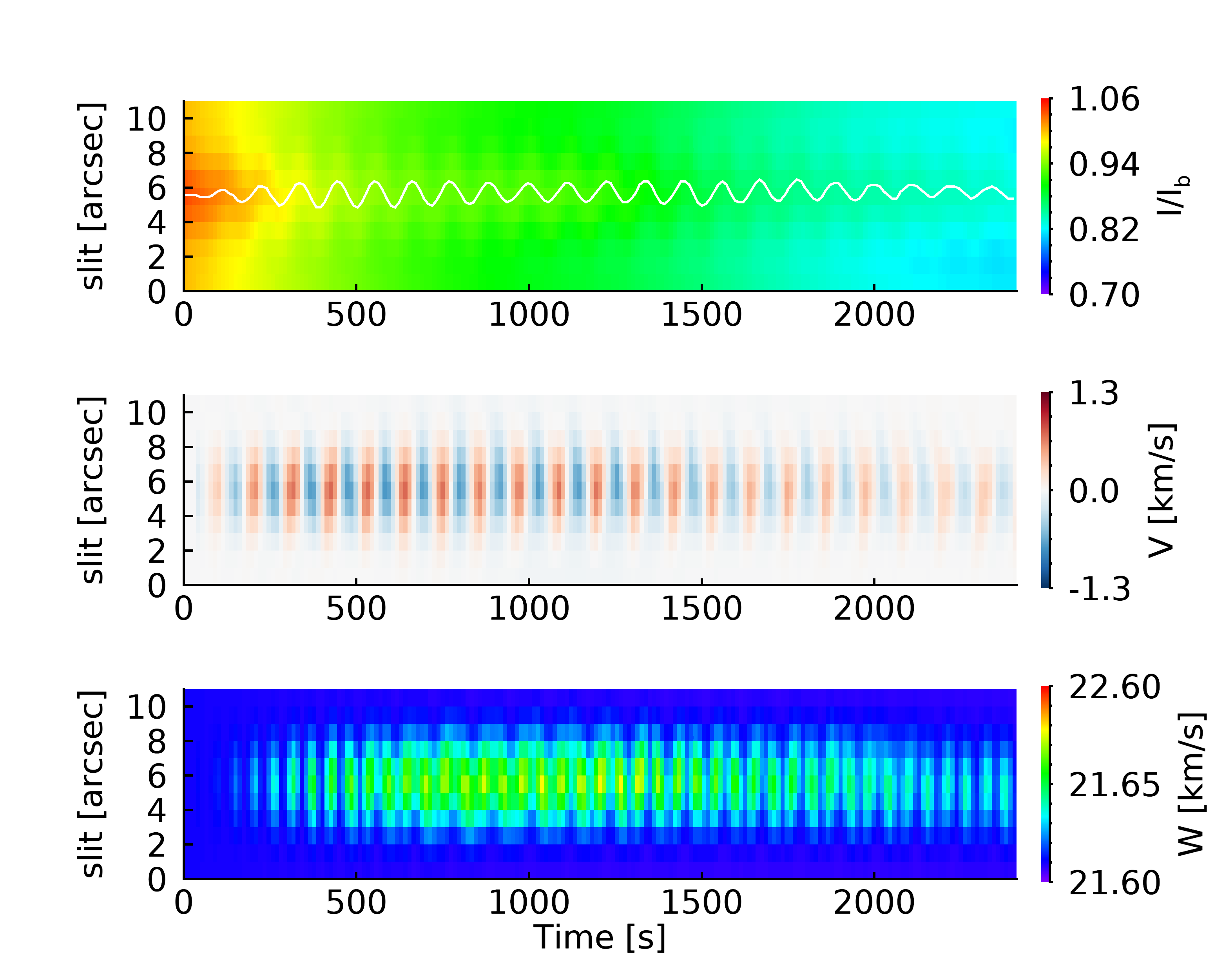}
		\includegraphics[width=0.5\textwidth,clip=]{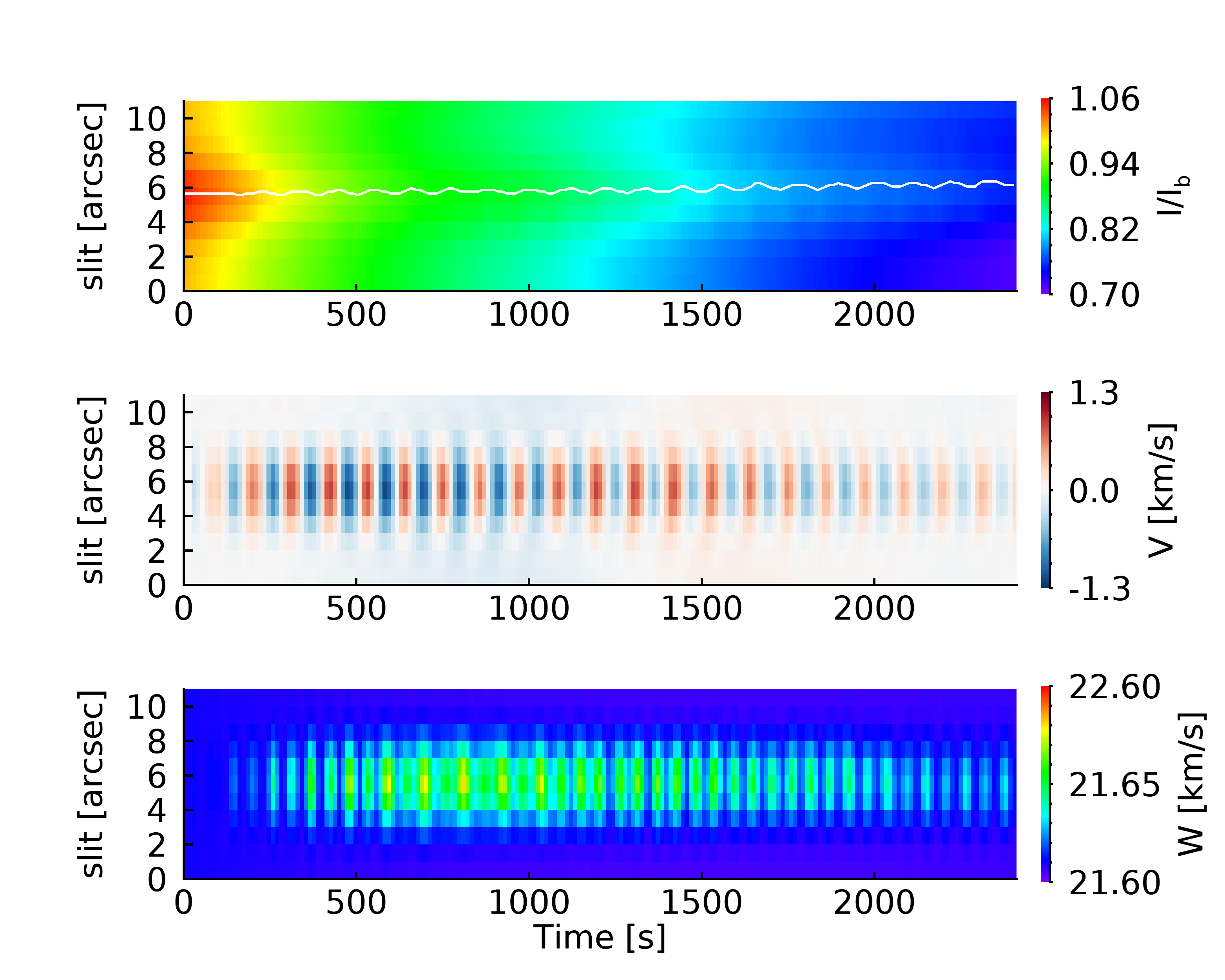}
	}
	\caption{Slit-time images of normalized intensity ($I/I_b$),
		Doppler velocity ($V$), and Doppler width ($W$)
		at slit 1 (left) and slit 2 (right) for Fe XII 195 \AA~line. 
		The white lines at the top panels are obtained by tracing the positions of peak intensity.}
	\label{F-spec_slit_195}
\end{figure}

\begin{figure}    %%%%%%%%%%%%%%%%%% FIGURE 8
	\centerline{\includegraphics[width=0.5\textwidth,clip=]{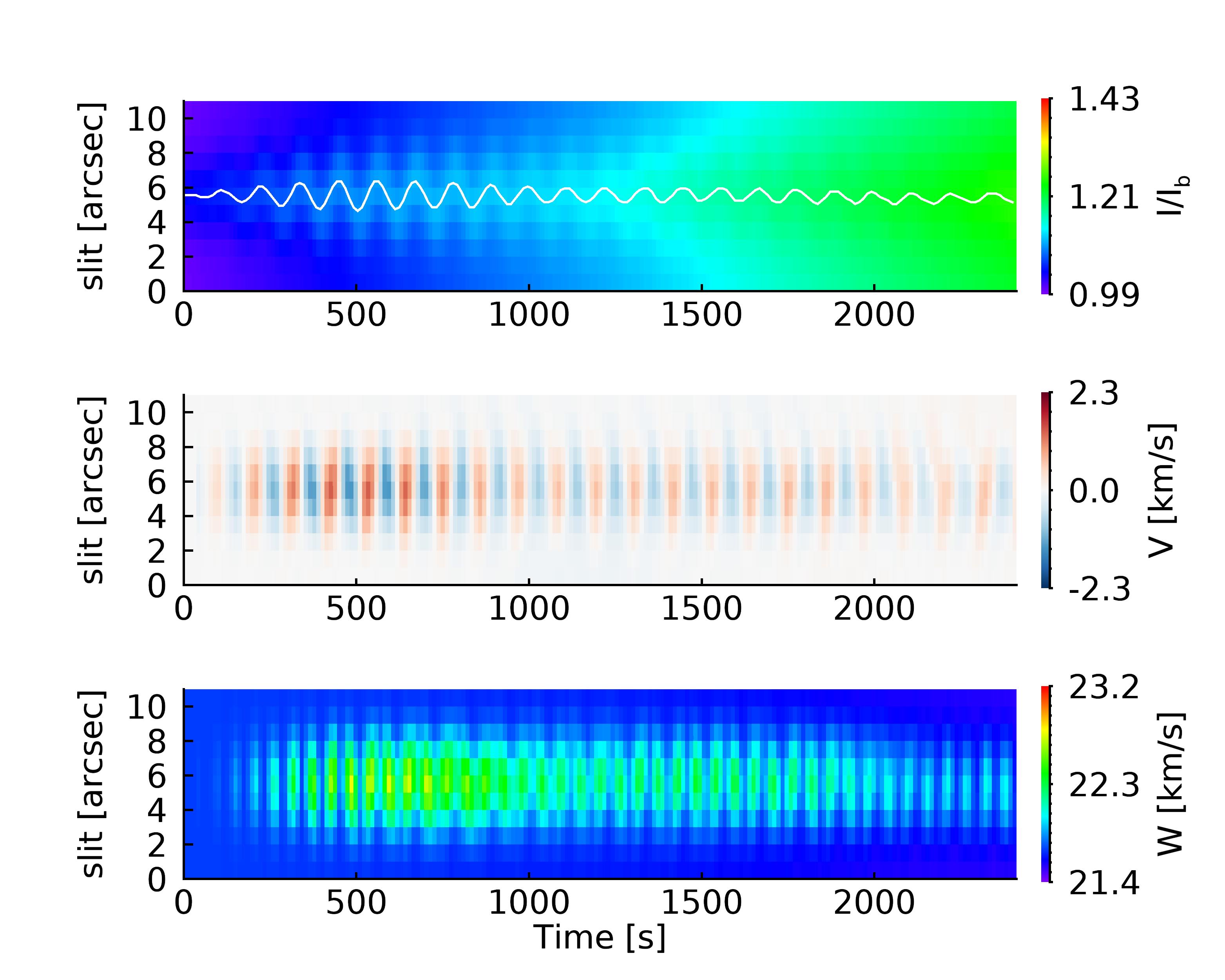}
		\includegraphics[width=0.5\textwidth,clip=]{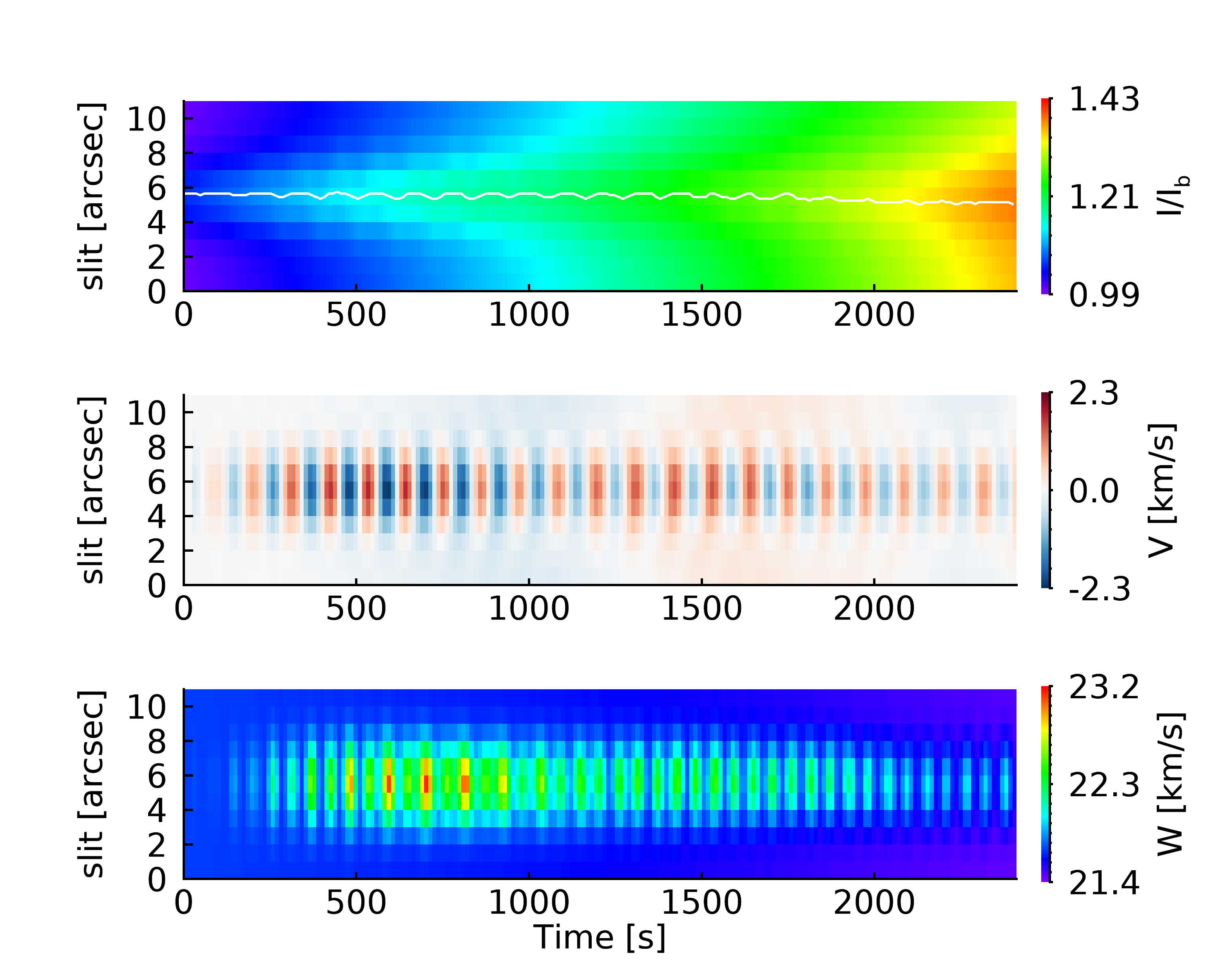}
	}
	\caption{Same as Figure \ref{F-spec_slit_195}, but for Fe VIII 185 \AA~line.}
	\label{F-spec_slit_185}
\end{figure}

\begin{figure}    %%%%%%%%%%%%%%%%%% FIGURE 9
	\centerline{\includegraphics[width=0.7\textwidth,clip=]{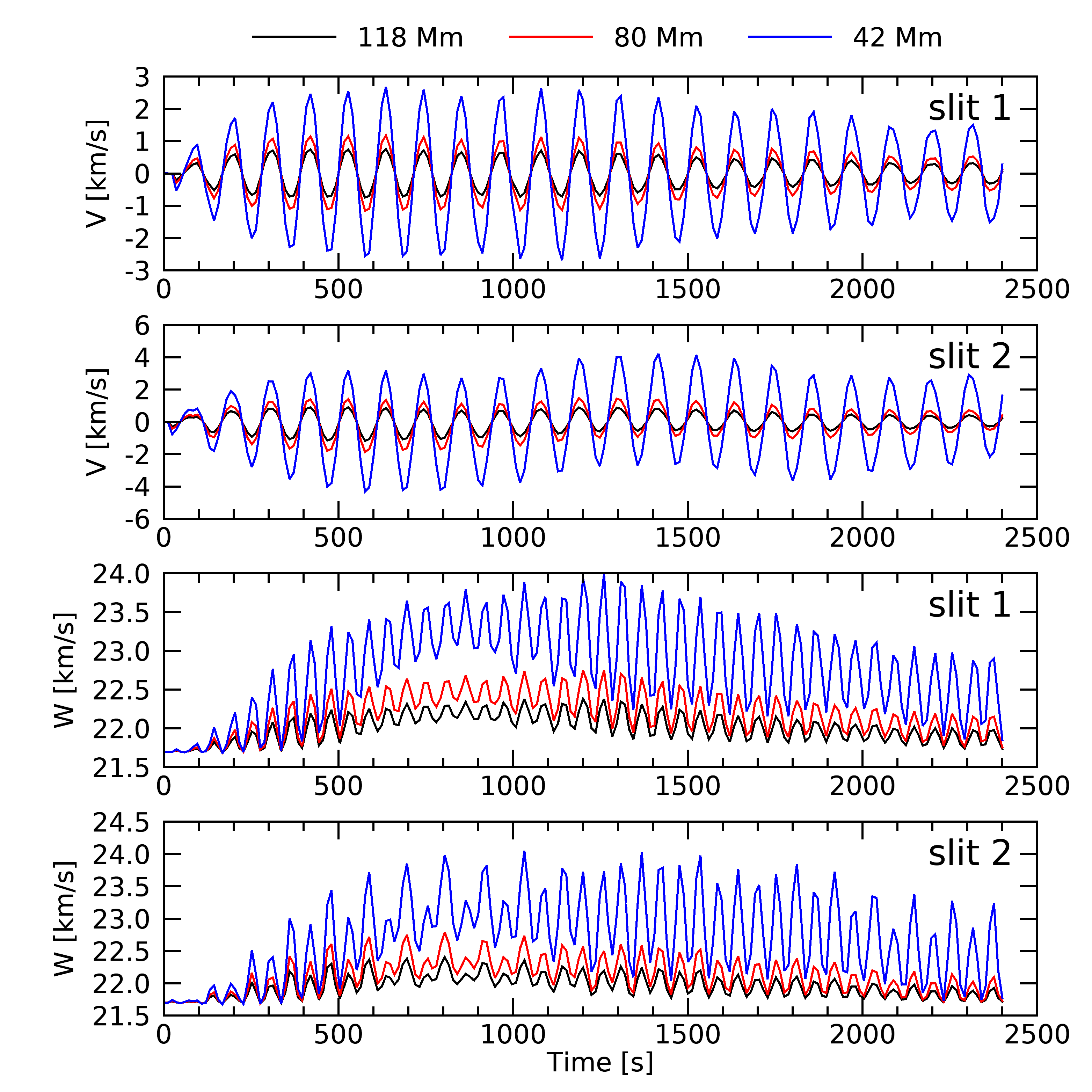}}
	\caption{Doppler velocity ($V$) and Doppler width ($W$) at slits centers of three different column depths for Fe XII 195 \AA~line.}
	\label{F-column_depth_195}
\end{figure}

\begin{figure}    %%%%%%%%%%%%%%%%%% FIGURE 10
	\centerline{\includegraphics[width=0.7\textwidth,clip=]{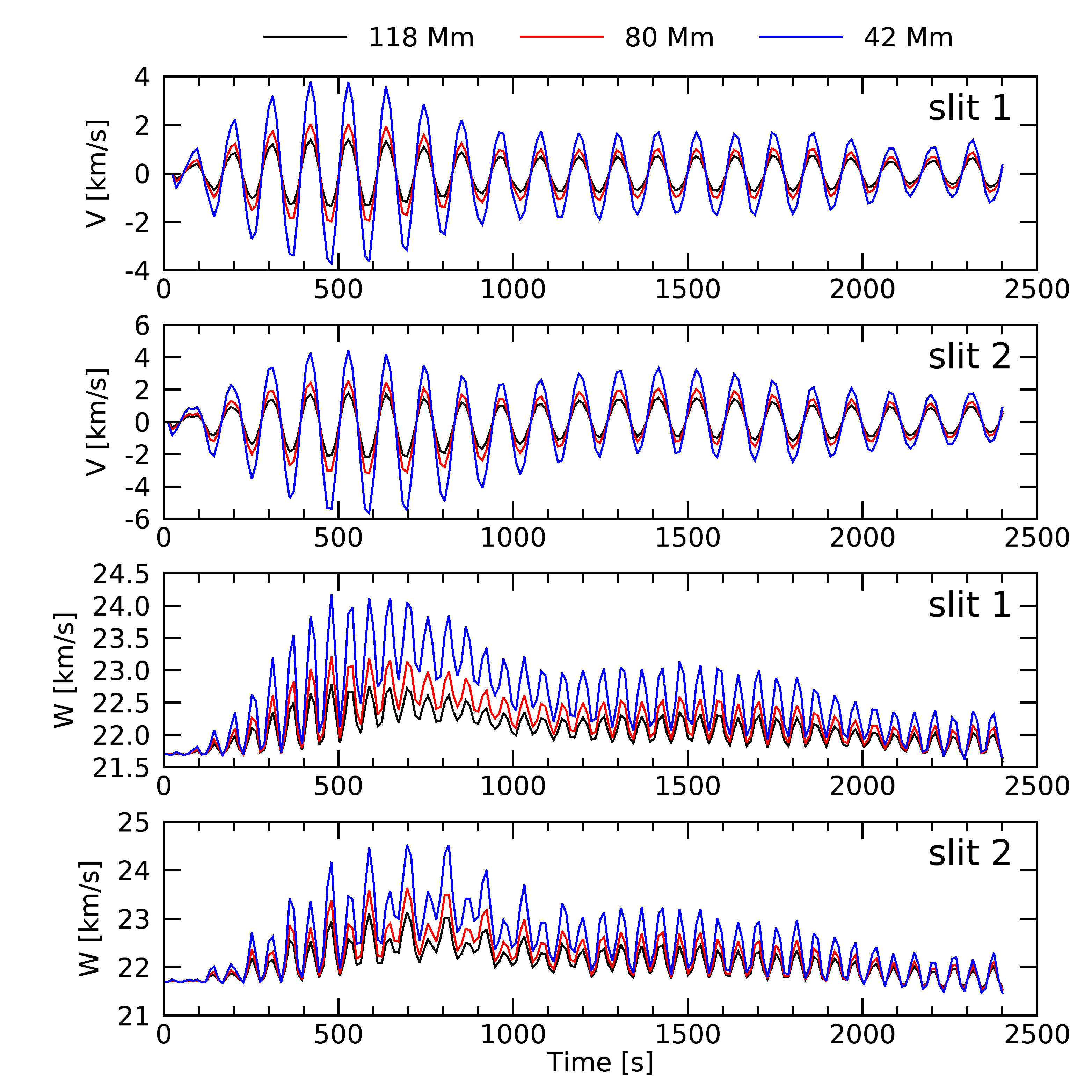}}
	\caption{Same as Figure \ref{F-column_depth_195}, but for Fe VIII 185\AA~line.}
	\label{F-column_depth_185}
\end{figure}

We find that even though the intensity changes for a different column depth,
the oscillation amplitude (white lines in Figures \ref{F-spec_slit_195} and \ref{F-spec_slit_185}) and velocity in the POS remain the same.
Thus the more important influence of the background emission 
is on the Doppler velocity and Doppler width. 
Figure \ref{F-column_depth_195} and Figure \ref{F-column_depth_185} show the Doppler velocity and Doppler width at the centers of 
both slits for Fe XII 195 \AA~line and Fe VIII 185 \AA~line, respectively.
We see that in the case of a small column depth,
both the Doppler velocity and Doppler width are obviously larger than in the case of a large column depth.
The oscillation amplitude of the Doppler velocity
increases first due to the development of kink oscillations.
At the later stage, the Doppler velocity decreases
 due to the development of small scale structures
which result in less coherent motions than at the earlier stage.
The rapid variations of the Doppler width is caused by periodic fluid motions of the kink oscillations.
Besides the rapid variations, the Doppler width has a trend of first increasing and later on decreasing.
The increasing trend is caused by resonant absorption and the development of KHI,
and decreasing trend is due to the radiative cooling
and the development of small scale structures.
%%Doppler width  platform in Figure 10 is probably due to the apex density concentration shown in Figure 3.

\begin{figure}    %%%%%%%%%%%%%%%%%% FIGURE 11
	\centerline{\includegraphics[width=\textwidth,clip=]{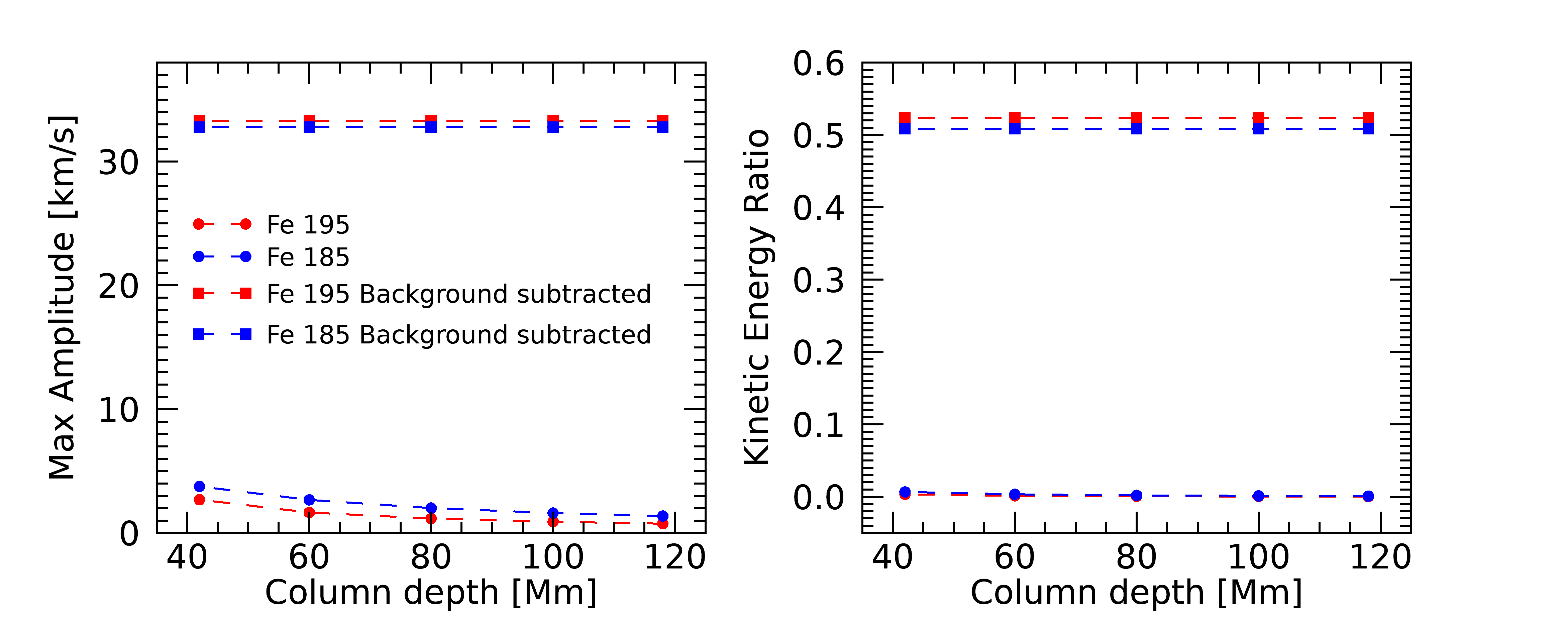}}
	\caption{Left: maximum oscillation amplitudes of the Doppler velocity for different column depths.
		Right: ratio of kinetic energy from the Doppler velocity with respect to the real kinetic energy (bottom panel of Figure \ref{F-kink_oscil}). The Doppler velocity is taken as the maximum value at the center of slit 1. The background subtracted value is got by fitting the subtracted monochromatic emissivity at the center of slit 1 by that at the edge of slit 1.}
	\label{F-amp_depth}
\end{figure}

The LOS Doppler velocity is of significance for estimating the kinetic energy of transverse waves in real observations.
However, our results show that the Doppler velocity
varies obviously for different column depths.
What's more, the unchanging background in our model significantly decreases the Doppler velocity.
Figure \ref{F-amp_depth} shows the maximum amplitude of Doppler velocity at the center of slit 1 and the corresponding kinetic energy ratio for different column depths.
In Figure \ref{F-amp_depth} we also show the background subtracted Doppler velocity,
which is got by fitting the subtracted
 monochromatic emissivity at the center of slit 1 
by that at the edge of slit 1.
We see that the Doppler velocity is obviously smaller for a larger column depth,
while the background subtracted Doppler velocity is roughly the same at different column depths.
More importantly, in our model, the Doppler velocity is one order of magnitude smaller than
both the background subtracted value and the real velocity in Figure \ref{F-kink_oscil},
similar as \cite{2017ApJ...836..219A}.
In the right panel of \ref{F-amp_depth}, we estimate the ratio of kinetic energy from Doppler velocity with respect to the real kinetic energy.
We assume a constant density and thus we simply calculate the ratio of velocity square with respect to the real one shown in the bottom panel of Figure \ref{F-kink_oscil}.
Our results show that the kinetic energy is two orders of magnitude smaller than the real value, 
while from the background subtracted Doppler velocity, the kinetic energy is around 50\% of the real kinetic energy.
The underestimation of the Doppler velocity and the associated kinetic energy was also found by \cite{2012ApJ...746...31D}
because of the integration through different strands.
The difference of our work is that we have only one single loop and the Doppler velocity is reduced because of the background emission.
Thus, it is
safe to speculate that the corona has both effects, 
and consequently,
the observed energy is much smaller than the real one.

We also estimate the kinetic energy using the POS velocity derived from the white lines in Figures \ref{F-spec_slit_195} and \ref{F-spec_slit_185}.
We find that the POS velocity is not influenced by the background emission,
and, for this particular LOS, 
the kinetic energy in POS is around 40\% to 50\% of the real value.
These results could 
increase the difficulty for reconstruction of 3D motions from POS and LOS motions proposed by \cite{2017ApJ...840...96S}.
Indeed, in that work, the LOS velocity was taken for granted, 
without taking into account the potential contribution of the background emission. 
Our results indicate that the observed velocity could be a severe underestimate of the real value. 
Thus, elliptical motions that
were found in that paper may have the wrong ellipticity or even orientation.
However, we can not reach a general conclusion on how much smaller the observed Doppler velocity is,
since besides the column depth,
the result would also depend on
the physical parameters inside and outside the loop 
(e.g., the density contrast, the temperature difference),
and could vary from case to case.

\subsection{Imaging Observations}
\begin{figure}    %%%%%%%%%%%%%%%%%% FIGURE 12
	\centerline{\includegraphics[width=0.7\textwidth,clip=]{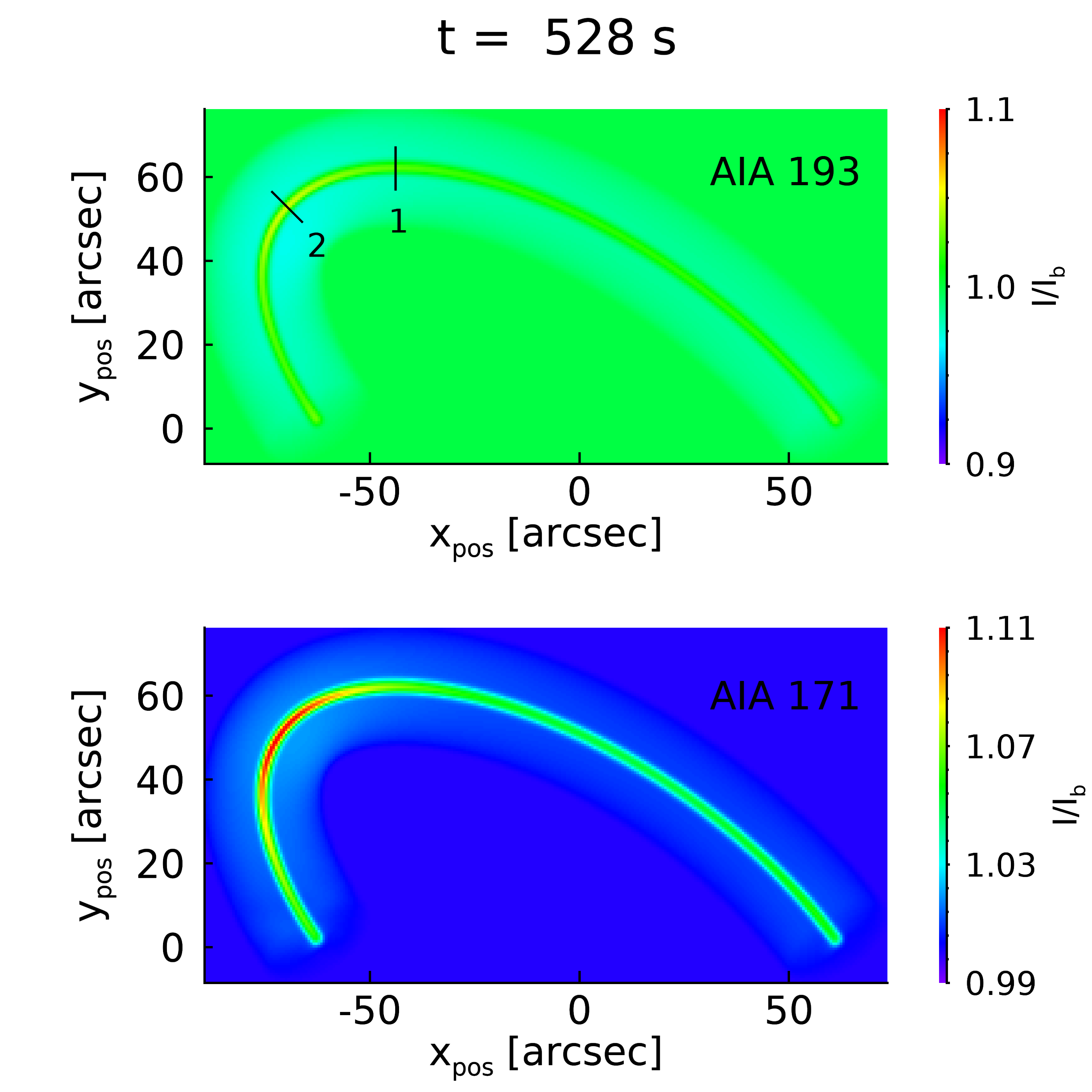}
	}
	\caption{Plane of sky images of normalized intensity ($I/I_b$),
		for AIA 193 (top) and AIA 171 (bottom) at $t = 528$ s.
		The black lines in the top panel mark the positions of two slits.
	An animated version of this figure is available. The animation has the same layout as the static figure, and runs from 0 -- 2400 s.
	\newline
	(An animation of this figure is available.)	
}
	\label{F-imaging}
\end{figure}

For imaging observations
we perform forward modeling of AIA 193 and AIA 171 channels,
in order to obtain the imaging observations in both hotter and cooler passbands.
Figure \ref{F-imaging} shows the normalized intensity in the plane of sky 
for AIA 193 (top) and AIA 171 (bottom) at one snapshot.
From Figure \ref{F-imaging} and the related animation,
we see that the loop oscillates as a whole 
and the small-scale structures can not be captured.
The intensity variation trends of AIA 193 and AIA 171 channels are opposite due to their different response functions.

\begin{figure}    %%%%%%%%%%%%%%%%%% FIGURE 13
	\centerline{\includegraphics[width=0.7\textwidth,clip=]{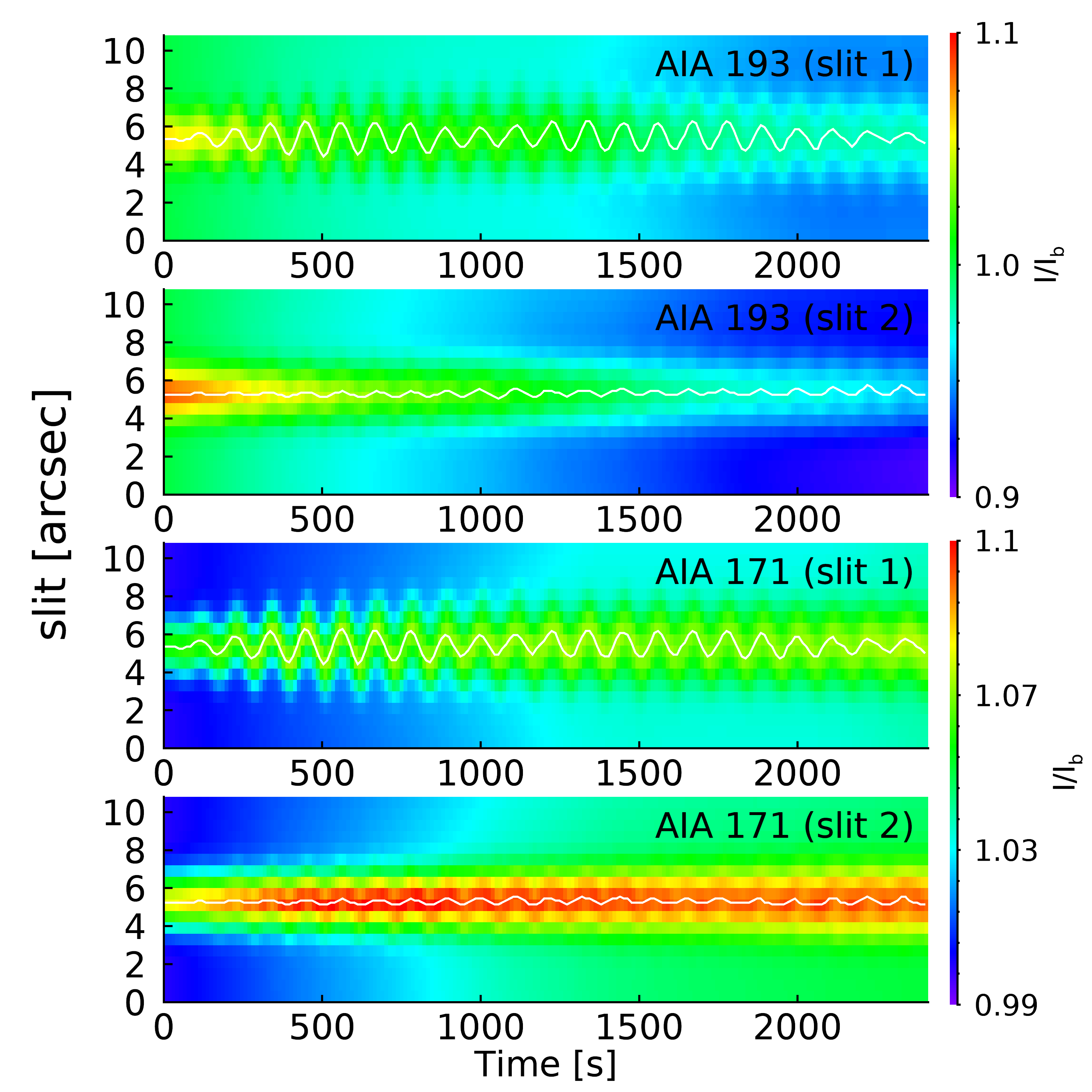}
	}
	\caption{Slit-time images of normalized intensity ($I/I_b$) for AIA 193 and AIA 171 at both slits.
		The white lines are obtained by tracing the positions of peak intensity.}
	\label{F-image_slit}
\end{figure}

X-t images for both channels and at both slits are shown
in Figure \ref{F-image_slit}.
At slit 2, the intensity oscillations are not obvious and the amplitudes are small,
while at slit 1 the periodic oscillations are clearly seen for both channels.
We also find that
at slit 1 the displacement amplitude and velocity in the POS
are similar to those of spectrometer observations,
and both of them are smaller than the kink amplitude and velocity shown in Figure \ref{F-kink_oscil}.

\section{Summary and Conclusion}
\label{S-summary}

In this work,
we simulate the transverse oscillations in a radiatively cooling coronal loop
that is being heated by transverse waves
and forward-model their EUV synthetic emissions.
The coronal loop is modeled as a straight, three-dimensional density enhanced magnetic cylinder, with gravity being ignored.
We drive the transverse oscillations using a dipole-like periodic
footpoint velocity driver.
A standing kink wave is formed and the Kelvin-Helmholtz instability develops and deforms the loop cross section to a turbulent state, 
resulting in localized energy dissipation and heating.
We also see a ponderomotive force induced long-period longitudinal flow which causes mass enhancement at the apex.

\begin{figure}    %%%%%%%%%%%%%%%%%% FIGURE 5
	\centerline{\includegraphics[width=0.6\textwidth,clip=]{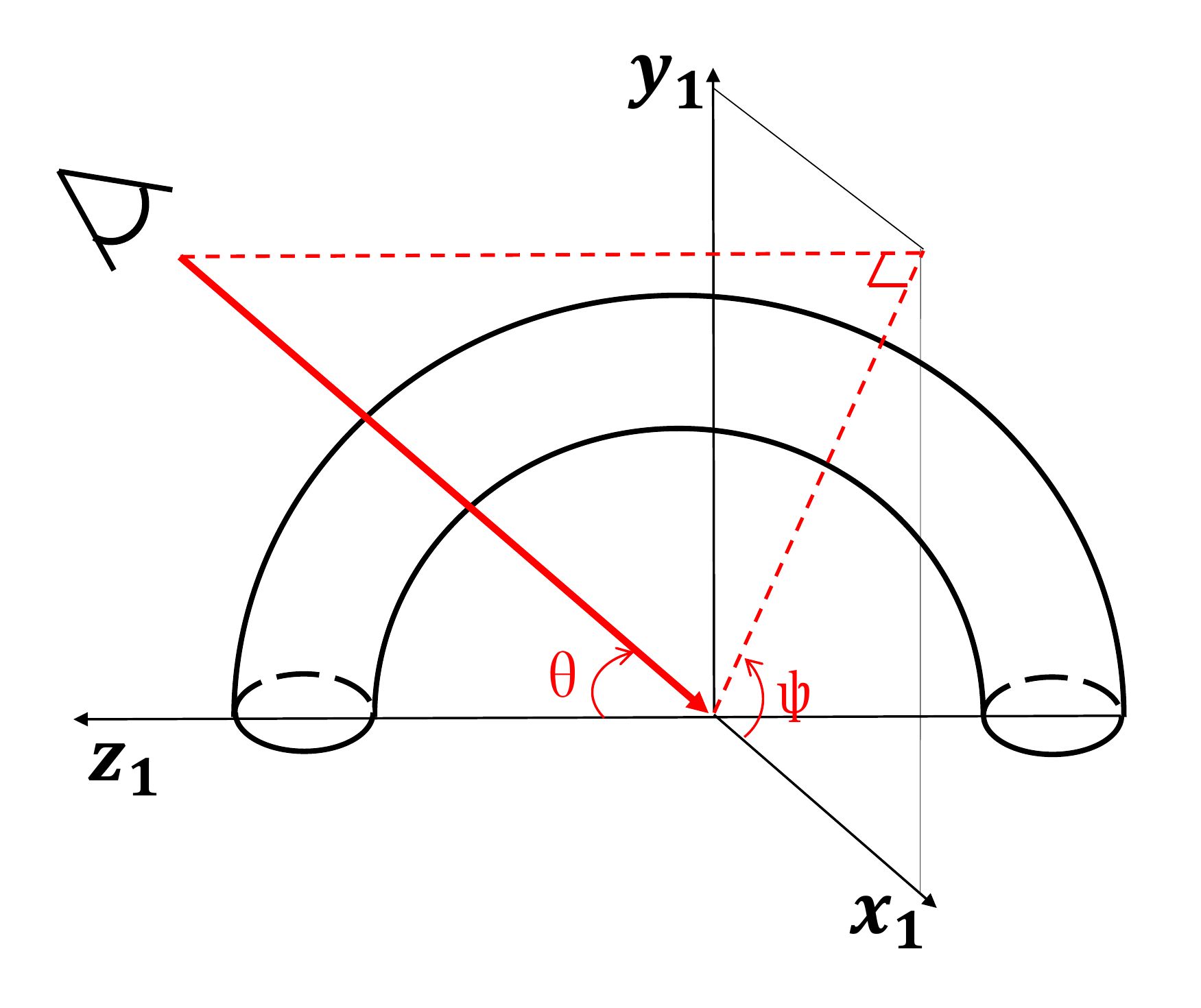}
	}
	\caption{Schematic of the semi-torus loop and line of sight (LOS) direction (red solid line, with $\theta=\psi = \pi/4$).}
	\label{F-los}
\end{figure}

We then bend the simulated straight loop to a semi-torus loop 
and forward-model the spectrometer and imaging observations at an oblique line of sight (Figure \ref{F-los}).
For spectrometer observations,
we forward-model emissions of Fe XII 195 \AA~and Fe VIII 185 \AA~lines,
and degrade the resolution to that of EIS.
For imaging observations,
we forward-model observations of AIA 193 and AIA 171 channels.
Our results show that the oscillation amplitudes of intensity
are different at different slits,
but are roughly the same for different spectral lines or channels.
The Doppler velocity shows periodic blue and red shifts in x-t images.
The Doppler width also shows periodic enhancement, with period being half of the kink oscillations,
matching the results for the impulsive kink mode investigated in \cite{2017ApJ...836..219A}. 
More importantly,
we find that both the Doppler velocity and Doppler width 
depends on the integration column depth.
The estimated kinetic energy from the LOS velocity is two orders of magnitude smaller than the real kinetic energy,
while from the POS velocity, the estimated energy is around 40\% to 50 \% of the real value.
These results are helpful for understanding transverse oscillations in coronal loops and their observational signatures,
as well as for estimating the wave energy in real observations.  

In our model, background subtraction can help recover the real oscillation velocities.
This result highlights the necessity of background subtraction in observations \cite[e.g.,][]{2003A&A...406.1089D,2017ApJ...842...38X} to get a more reliable velocity.
In our idealized model, the POS oscillation and velocity are not influenced by the background emission.
However, the results could be different for real observations where noise plays an additional role.
The noise effects are not studied in this work.

It is also worth noting that the low amplitude oscillations (slit 2 in our model) are not well measured
due to the limitation of spatial resolution.
The upcoming Multi-slit Solar Explorer
\citep[MUSE,][]{2020ApJ...888....3D}
mission, with both higher spatial resolution and temporal cadence,
could perform better observations of these low amplitude oscillations.
Moreover, we have shown that the background emission plays a major role
in determining the observed Doppler velocity. 
Even with significant driving
amplitudes, 
our Doppler velocities only go up to 1km/s for deep LOSs,
which puts it at the detectability limit of the proposed spectral resolution of MUSE.

\begin{figure}    %%%%%%%%%%%%%%%%%% FIGURE 14
	\centerline{\includegraphics[width=0.7\textwidth,clip=]{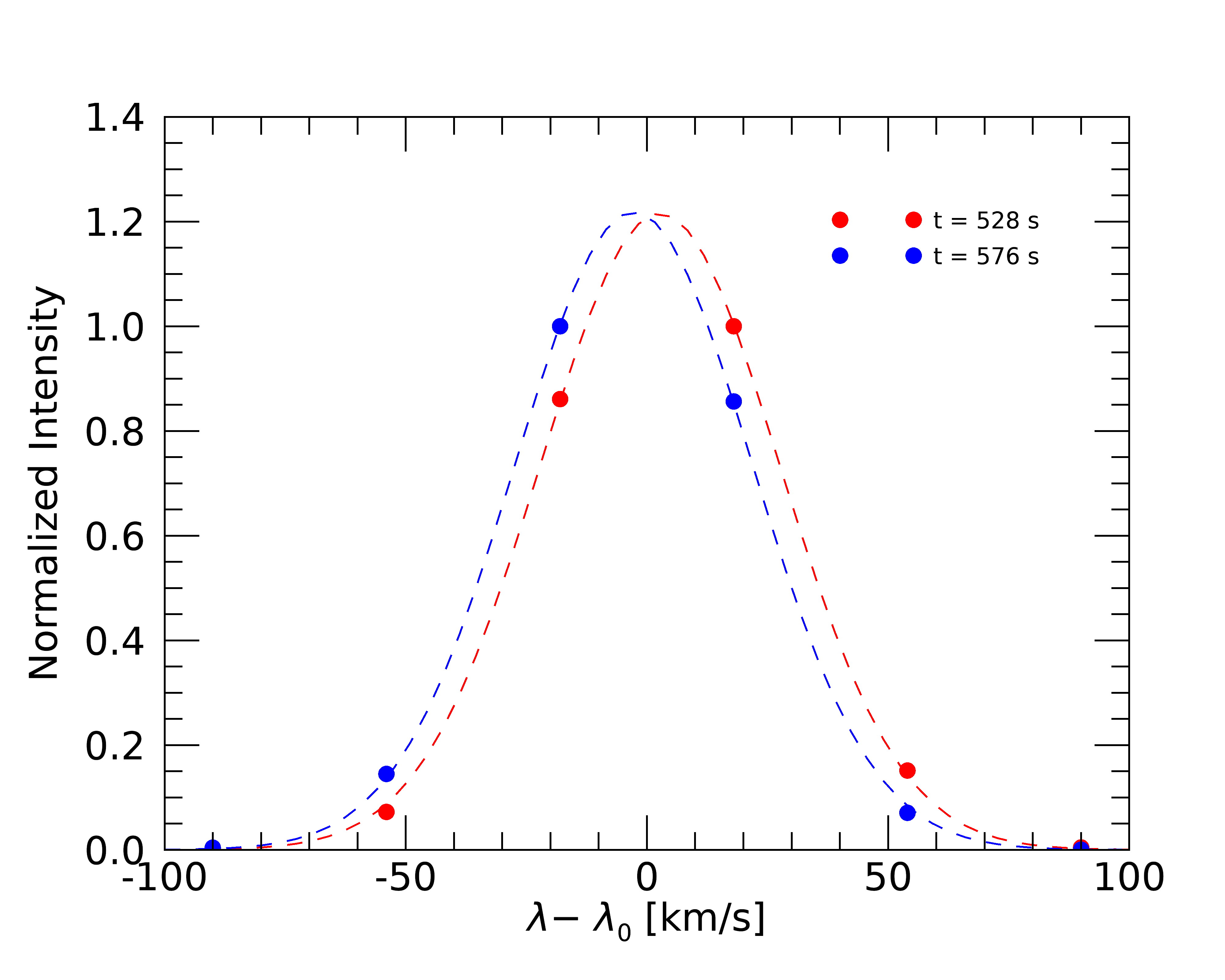}
	}
	\caption{Spectral profile and the single gaussian fit at the center of slit 1 in the case of Fe XII 195 \AA~line and the 42Mm column depth.}
	\label{F-spec_profile}
\end{figure}

In our analysis, a single gaussian fit is used to derive the Doppler velocity and Doppler width.
Even though the single gaussian fit is a common practice,
in some observations, double or multiple gaussians are used
\citep[e.g.,][]{2012ApJ...759..144T,2015ApJ...810...46H,2015ApJ...807...71P}.
Double or multiple gaussian profile is usually formed due to the blending of different spectral lines or the flows that have significant contribution to the emission profile.
However, in our model, we find that it is justifiable to carry out the single gaussian fit.
Figure \ref{F-spec_profile}
shows the degraded monochromatic intensity at the center of slit 1 and the corresponding single gaussian best fit (dashed lines) at two snapshots in the case of 42 Mm column depth.
Due to the low spectral resolution of EIS ($\approx$36 km/s),
the degraded spectral profile does not show enough information or clear evidence that double or multiple gaussians should be carried out for the fitting.
Thus the single gaussian fit is reasonable in our model.
	
Our current model has a number
of inherent limitations,
and the scenario in real observations is much more complex than in our model.
Our model is still a very simplified one, as we do not include some physical processes such as gravity and thermal conduction.
The curved loop is achieved by simply bending the simulated straight loop.
A more realistic model, including, e.g., density stratification, curved loop, and thermal conduction is planned in the future, in order to have a direct comparison with the real observations.

\acknowledgments
We thank the referee for the comments that help improve this manuscript.
This work is supported by the National Natural Science Foundation of China (41904150, 11761141002, 41974200) 
and the European Research Council (ERC) under the European Union’s Horizon 2020 research and innovation
program (grant agreement No. 724326).
P.A. acknowledges funding from his STFC Ernest Rutherford Fellowship (No. ST/R004285/2).
M.S. is supported by the international postdoctoral exchange fellowship from China Postdoctoral Council.
CHIANTI is a collaborative project involving the University of Cambridge (UK), the NASA Goddard Space Flight Center (USA), the George Mason University (GMU, USA) and the University of Michigan (USA).
The computational resources and services used in this work were provided by the VSC (Flemish Supercomputer Center), funded by the Research Foundation Flanders (FWO) and the Flemish Government-department EWI.

\bibliographystyle{apj}
\bibliography{export-bibtex}


\clearpage

\clearpage

\clearpage

\clearpage

\clearpage

\clearpage

\clearpage

\clearpage

\clearpage

\end{document}